\documentstyle[epsf,12pt]{article}
\catcode`\@=11
\def\numberbysection{\@addtoreset{equation}{section}
  \def\theequation{\arabic{section}.\arabic{equation}}}
\begin{document}
\addtolength{\baselineskip}{3.0mm}
\begin{titlepage}
%  \vspace{5mm}
\begin{flushright}{
    TIT/HEP-272/COSMO-49\\ February, 1995}
\end{flushright}
%\vspace{2mm}
\begin{center}
  {\Large\bf Dynamics of Yang-Mills Cosmology Bubbles in Bartnik-McKinnon
Spacetimes}\\
\vspace{4mm}
{\bf Shuxue
Ding}\def\thefootnote{\fnsymbol{footnote}}\footnote{Electronic address:
sding@th.phys.titech.ac.jp}

\vspace{2mm}
{ Department of Physics, Tokyo Institute of
    Technology,\\ Oh-Okayama, Meguro, Tokyo 152, Japan\\}

\vspace{8mm}
{\large ABSTRACT}
\end{center}
We investigate the dynamics of a Yang-Mills cosmology (YMC, the FRW
type spacetime) bubble in the Bartnik-McKinnon (BK) spacetimes.
Because a BK spacetime can be identified to a YMC spacetime with a
finite scale factor in the neighborhood of the origin, we can give a
natural initial condition for the YMC bubble. The YMC bubble can
smoothly emerge from the origin without an initial singularity.
Under a certain condition, the bubble develops continuously
and finally replaces the entire BK spacetime, the metric of which
is the same as the one of the radiation dominated universe. We
also discuss why an initial singularity can be avoided in the
present case in spite of the singularity theorems by Hawking
and Penrose.

\hfill\break
\noindent
PACS number(s): 98.80.Hw, 04.20.Dw, 98.80.Cq.
\end{titlepage}
\clearpage \setcounter{section}{0}
\setcounter{equation}{0}
%%%%%%%%%%%%%%%%%%%%%%%%%%%%%%%%%%%%%%%%%%%%%%%%%%%%%%%%%%%%%%%%%%%%%%%%%%%%%%%
%%%%%%%%%%%%     Section 1. Introduction                     %%%%%%%%%%%%%%%%%%
%%%%%%%%%%%%%%%%%%%%%%%%%%%%%%%%%%%%%%%%%%%%%%%%%%%%%%%%%%%%%%%%%%%%%%%%%%%%%%%
\numberbysection
\section{Introduction}

A remarkable pioneering work of Israel was writing the Einstein
field equation into the form of junction conditions at the
boundary between two spacetime regions when the metric is
continuous but the curvature and the energy-momentum tensor
are not, having some form of singularity \cite{Israel,MTW}.
An example is a bubble with a domain wall or a thin shell,
on which the singularity of the energy-momentum is
of the form of the $\delta$ function, and is allowed without
violation of the Einstein field equation. How bubbles develop
is also an interesting topic in the context of real cosmology
\cite{VKO,Col1,CD,SSKM,BKT1,ADLS,BGG,FG,AKMS,BKT2,Guth,PV,BV,ABV,V}.
To the best knowledge of the author, there are three reasons for
such cosmological interests.

The first is the discussion about child universes. A child
universe can develop from a small region by an inflation. If
this inflation is due to a ``false vacuum'' of scalar field
which is equivalent to a positive cosmological constant, the
false vacuum region is separated from the true vacuum region by
a domain wall. The motion of this domain wall just obeys the
junction conditions.

The second is the inhomogeneity by local inflations. We can
imagine that in the initial universe, there are not only one
of the regions entered into the ``false vacuum'' states of
the scalar field. Because such regions will undergo various
magnitude of inflations, comparing to the regions which have
not undergone any inflation, they will have bigger size and
mass densities. Since in different regions, the potential
heights of the metastable states are not necessarily equal,
the degrees of inflations are also different. Then the
development of the inhomogeneity by the phase transitions
can be investigated by the consideration of the dynamics of
the bubbles.

The third is most intriguing. As the existence of the ``false
vacuum'' of a scalar field is sufficient for an inflation,
Farhi and Guth discussed \cite{FG,Guth} the possibility of
realization of an inflation in the present universe, especially
by {\it in principle} man-made processes. They especially
considered a spherically symmetric de Sitter bubble created in,
necessarily by Birkhoff theorem, the Schwarzschild spacetime.
Since a spherically symmetric bubble should be created
classically at $r = 0$, the origin of the Schwarzschild
coordinates, where a black hole singularity is located, we
always have it as an initial singularity of a bubble.
Classically, an initial singularity becomes an obstacle
because human being does not know how to prepare an initial
singularity in the present universe. Such a singularity may be
avoided by a quantum tunneling \cite{Guth}. However, no one
could give a complete discussion by an explicit example up
to now.

To pursue the idea of Farhi and Guth further is one of the
purposes of this paper. We will reconsider the bubble creation
in classical sense and examine the possibility to avoid the
initial singularity \cite{In}. Let us first analyze the initial
singularity which necessarily exists in the de Sitter --
Schwarzschild bubble. At first, we find that a de Sitter
spacetime expands so fast that such a bubble can
always contain an anti-trapped surface \cite{FG}. Secondly,
the parent spacetime -- a Schwarzschild spacetime has a
singularity at $r = 0$, where the spherically symmetric bubble
emerges. For the present purpose, we should consider a bubble in which
the spacetime expands less fast that it does not contain an
anti-trapped surface in the bubble developing region, and a
parent spacetime which is regular everywhere. For the first condition,
we may consider a Friedmann-Robertson-Walker (FRW) bubble. Although
there also exist anti-trapped surfaces in a region of the pure FRW
spacetime, we shall see in the present paper that the bubble with a
non-spacelike shell cannot enter the region to contain an anti-trapped
surface. However, if the source of the spacetime is a usual ideal
liquid, by Birkhoff theorem, the exterior of the bubble is inevitably
a part of the Schwarzschild spacetime too. Then the second condition
cannot be satisfied. However, if there exists a long
range field in the FRW bubble which can penetrate the
shell, we can be free from Birkhoff theorem since the outside
is not a vacuum, and have a parent spacetime which is regular
everywhere. Previously Yang-Mills cosmology (YMC) solutions were
found and carefully discussed \cite{GV1,GS2,H,VD,Ding}. These
spacetimes are homogeneous and isotropic, and equivalent to FRW
spacetimes with radiation dominant source \cite{GV1}. Because we,
for simplicity, only consider a spherically symmetric bubble, the
spacetime outside of the bubble should be static and spherically
symmetric, and should be one of the following possibilities:
Bartnik-McKinnon (BK) particle-like solutions
\cite{BMcK,VG,SZ1,MW,GS1} or Bizon - Kunzle - Masood-ul-Alam black
holes \cite{Bizon}, depending on the initial spacetime in which a
bubble is going to be created. For the case of a Bizon - Kuenzle -
Masood-ul-Alam black hole, the situation is similar to that of
Schwarzschild spacetime. Thus we shall consider a YMC bubble
developing in a Bartnik-McKinnon spacetime. The BK spacetime, which is
regular everywhere, is just what we need. We expect that a YMC-BK
bubble can be created without an initial singularity.

Homogeneous and isotropic solutions of coupled system of gravity and
Yang-Mills field were discussed enormously
\cite{GV1,GS2,H,VD,Ding}. These discussions concerned both of
Euclidean (whormhole) and Lorentzian signature solutions. The YMC
\cite{GV1} are Lorentzian ones which are equivalent to all types (closed,
open and spatially flat) of FRW spacetimes in the radiation dominant
epoch. The solutions are also called hot universes with cold matters.
For the convenience of use, we shall give a brief review for the
solutions in the subsection {\it a} of the section 2.

The Bartnik-McKinnon solutions \cite{BMcK} are asymptotically
flat solutions of coupled system of gravity and Yang-Mills field,
which are the spherically symmetric. The solution in the whole
spacetime is regular. An important fact is that this spacetime in
the neighborhood of the origin can be set to be the same as a YMC
(FRW) spacetime due to the equality of the curvatures there. For a
YMC bubble created from BK spacetime, if there is no initial
singularity, a proper initial condition can/should be given by the
equivalence. Then we can imagine a bubble emerging and developing in
this spacetime. For the purpose of
convenience of the discussion, we shall give a brief review
for this kind of solution in the subsection {\it b} of the
section 2, there we also give some other new facts and further
analysis, especially global ones, for our purposes.

Although we have a static, spherically symmetric spacetime
-- the BK spacetime -- as an initial spacetime of a YMC bubble
without initial singularity, we still could not confirm whether
such a bubble can develop in a BK spacetime and finally becomes a
child universe or not before our investigation of the equation
of motion for the shell. In the section 3, we shall discuss how
a YMC bubble develops in a BK spacetime quantitatively. There exists
a shell trajectory between the two spacetime regions. We shall
discuss the equation of motion for the bubble and examine whether the
bubble can expand to the extent comparable to the corresponding
FRW universe or not. To discuss a shell in an exact sense, we
should discuss a field configuration which asymptotically
approaches to two different regions, the YMC and BK spacetimes.
Investigating numerically how a BK solution evolves using the
perturbation theory, Zhou and Straumann \cite{ZS} found some
shell-like structure with a finite thickness. However, to discuss
such an exact shell is very complicated and is impossible
analytically. Numerical methods are not effective for the
discussions of global properties and the final
stage of the bubble. Because we are interested in the cosmological
case, the thin shell approximation is always sufficient for it (e.g,
see \cite{SSKM,BKT1,BGG}). That is to say, when the shell is very thin
comparing to the any other length of interest, the junction condition
method is good enough to solve the Einstein field equation with a
curvature jump. In the section 4, we found, under some conditions, a
open or spatially flat type of bubble can expand infinitely and
finally replace the entire BK spacetime.

In fact, from only the viewpoint of a shell between YMC
and BK spacetimes, there are some interesting properties which
are different from those of de Sitter-Schwarzschild bubble.
Since the pressure inside the bubble is always higher than
that outside of the shell, except at the $\rho = 0 $ where an
initial condition makes them equal, the bubble really
continuously emerges and pushes into the parent spacetime.
Of course it also expands to the direction of the YMC spacetime.

Hawking showed that there is an initial singularity in the
standard big bang cosmology model \cite{HE}. We should answer
why we can create a FRW type bubble without an initial singularity.
One might argue such a bubble spacetime is not a maximally
extended spacetime to which the singularity theorems \cite{HP,W}
apply. However the total spacetime composed of bubble, outside of
the bubble in the parent spacetime and the shell can be regarded as
a maximally extended spacetime. In fact, given the energy-momentum
tensor in both bubble and parent spacetimes, together with
it on the shell, one can always solve the Einstein field
equation to get a maximally extended spacetime. In the paper of
Farhi and Guth, they already implicated this fact \cite{FG}.
Then what is the reason that we can avoid an initial singularity?
After a detailed analysis, we find that there is fundamental
difference between a universe created from nothing and a bubble
emerging and developing inside another spacetime. Since the universe
expands, spatial hypersurfaces may contain anti-trapped surfaces
\cite{HE}. However for the case of a bubble inside the other
spacetime, if the bubble is a FRW type spacetime, the bubble only
corresponds to a part of a spatial hypersurface in the FRW
spacetime. Thus a bubble may or may not contain an anti-trapped
surface. In the section 5, we shall show that a non-spacelike shell,
as we expected, does not contain an anti-trapped surface and give other
detailed discussion about singularity theorems. To make a comparison,
we shall also discuss why there exists an initial singularity for a de
Sitter-Schwarzschild bubble.

In the section 6, we draw our conclusion and give some discussions.

\vspace{5mm}
%%%%%%%%%%%%%%%%%%%%%%%%%%%%%%%%%%%%%%%%%%%%%%%%%%%%%%%%%%%%%%%%%%%%%%%%%%%%%%%
%%%%%%% Section 2. Bartnik-McKinnon and Yang-Mills Cosmology solutions%%%%%%%%
%%%%%%%%%%%%%%%%%%%%%%%%%%%%%%%%%%%%%%%%%%%%%%%%%%%%%%%%%%%%%%%%%%%%%%%%%%%%%%%

\section{Yang-Mills Cosmology and Bartnik-McKinnon Solutions -- Reviews}

\noindent
{\it a. Yang-Mills Cosmology (YMC)}
\vspace{5mm}

Yang-Mills Cosmologies (YMCs) \cite{H,VD,GS2} are spacetimes
which are homogeneous and isotropic with the Yang-Mills field
as a source and are the same as the FRW universes with radiation
dominant perfect liquid.

The metric for the spacetime is assumed to be the form of
FRW universe,
\begin{equation}
ds^2 = - dt^2 + a^2(t)(d{\chi}^2 + f^2(\chi){d\Omega}^2) =
a^2(\lambda) ( -d{\lambda}^2 + d{\chi}^2 + f^2(\chi){d\Omega}^2),
\label{frw}
\end{equation}
where $f(\chi) = sin(\chi), \ sinh(\chi)$, and $\chi$ for closed ($k =
1$), open ($k = -1$), and spatially flat ($k = 0$) Friedmann models,
respectively. Here $d{\Omega}^2 = d{\theta}^2 + sin^2(\theta)
d{\phi}^2$, and  $\lambda $ is the conformal time which relates
to the cosmic time $t$ by $d\lambda/dt = a^{-1}$.

By the Einstein field equation, the line element (\ref{frw}) defines
for the energy-momentum tensor $T_{\mu \nu}$ the typical perfect fluid
structure:
\begin{equation}
T^0_0 = - \epsilon (\lambda), \ \ \ T^i_k = p(\lambda){\delta}^i_k.
\end{equation}
Since the Yang-Mills field is conformal invariant, one has $p =
\epsilon/3$. Thus the Yang-Mills cosmology is always radiation
dominant. Together with the Bianchi identities, the energy-momentum
tensor has to take the form
\begin{equation}
T^{\mu}_{\nu} = {C^2 \over 2e^2 a^4}diag(-3, 1, 1, 1),
\end{equation}
where $C^2$ is some constant, and $e$ is the coupling constant of the
Yang-Mill field. The Friedmann equation becomes
\begin{equation}
({da \over d\lambda})^2 + k a^2 = {2 \kappa C^2 \over e^2},
\end{equation}
where $k = 1, -1, 0$ for closed, open and spatially flat Friedmann
models, respectively. In this paper, $\kappa = 8\pi G $, where $G$
denotes the gravitational constant.

The line element (\ref{frw}) is invariant under actions of the spatial
rotations and translations. Thus the gauge field symmetry must be the
same. For simplicities, we shall discuss the gauge field with only spatial
rotation symmetry at first and then pick out the solution which is also
invariant under actions of translations.
The well-known Witten ansatz is suitable for our purpose
 \begin{equation}
e{\bf A} ={\omega}_0 L_1 d\lambda + {\omega}_1 L_1 dr - g(L_3
d\theta - L_2 sin\theta d\phi),
\label{PO}
\end{equation}
where $f$ and ${\omega}_i$ are functions of $\lambda$ and $\chi$. Here
a coordinate dependent gauge group generators $L_a$, which satisfy the
commutative relation $[L_a, L_a] = i{\epsilon}_{abc} L_c$, are
introduced:
\[
L_1 = sin\theta cos\phi {{\sigma}_1 \over 2} + sin\theta sin\phi
{{\sigma}_2 \over 2} + cos\theta {{\sigma}_3 \over 2},
\]
\begin{equation}
L_2 = {\partial}_{\theta}L_1, \ \ \ \  L_3 = {1 \over sin\theta}
{\partial}_{\phi}L_1,
\end{equation}
where ${\sigma}_i$ are the Pauli matrices.

{}From the Einstein and the Yang-Mills field equations, we obtain
\begin{equation}
{\omega}_0 = -{f \sqrt{1 - kf^2} \over 1 + f^2({\psi}^2
- k)}{d\psi \over d\lambda}, \ \ \ {\omega}_1 = {{\psi}f^2
  ({\psi}^2 -1) \over {1 + f^2({\psi}^2 - k)}},
\label{omega}
\end{equation}
and
\begin{equation}
g =1 - \sqrt{1 + f^2({\psi}^2 - k)}.
\label{f}
\end{equation}
Here $\psi$ satisfies the equation
\begin{equation}
({d\psi \over d\lambda})^2 + ({\psi}^2 - k)^2 = C^2.
\end{equation}

By checking directly, we can find that the Yang-Mills field with
${\omega}_i$ and $f$ given by the equations (\ref{omega}) and
(\ref{f}) give a homogeneous energy-momentum tensor as the source
of the FRW type of spacetime. That is to say, The solutions are also
invariant under the action of translations.

\vspace{5mm}
\noindent
{\it b. Bartnik-McKinnon (BK) Solutions}
\vspace{5mm}

The static spherically symmetric (SSS) solutions of
Einstein-Yang-Mills (EYM) equations have been paid much
attention since the last decade after an explicit series of
such solutions was found by Bartnik and McKinnon
\cite{BMcK,VG,SZ1,MW,GS1}. Before that the existence of such
kind of solutions was doubted because neither the vacuum Einstein
nor the pure Yang-Mills equations have non-trivial regular
SSS solutions with finite energy and there is no such a solution
for EYM equation in (2+1)-dimensions.

All of the solutions found by Bartnik and McKinnon are
numerical. What is interesting is that the solutions were shown
to be analogues of sphalerons in the pure Yang-Mills theory or
the Weinberg-Salam model \cite{VG,Bizon,SZ1,MW,GS1}, which may
be called EYM sphalerons.

The Yang-Mills connection 1-form ansatz for the solutions is
\begin{equation}
e{\bf A} = w{\sigma}_1 d{\theta} + (cot {\theta} {\sigma}_3 +
w{\sigma}_2) sin {\theta} d{\phi}, \\
\label{pot}
\end{equation}
where $w$ is an unknown function which depends only on $r$.
The metric is
\begin{equation}
ds^2 = -T^{-2} dt^2 + R^2 dr^2 + r^2 d {\Omega}^2,\\
\label{met}
\end{equation}
where $T$ and $R$ are some functions of $r$.

Rescaling the variable $r \rightarrow {\sqrt{\kappa} \over e}r$
removes the dependence on $\kappa$ and $e$ from the Einstein and
Yang-Mills (EYM) equations. Introducing a convenient variable $m(r)$
by writing $R = ( 1 - 2 m(r)/r)^{-1/2}$, the static spherically
symmetric EYM equations read
\begin{equation}
m^{\prime} = ( 1 - {2m \over r}) w^{\prime 2} + {1 \over 2 }{(1
  - w^2)^2 \over r^2},
\label{ein1}
\end{equation}
\begin{equation}
r^2(1 - {2m \over r})w^{\prime \prime} + [2m - {(1
  - w^2)^2 \over r}]w^{\prime} + (1 - w^2)w = 0,
\label{ya1}
\end{equation}
\begin{equation}
2r( 1 - {2m \over r}) {T^{\prime} \over T} = {(1
  - w^2)^2 \over r^2} - 2( 1 - {2m \over r})w^{\prime 2} - {2m
  \over r},
\label{ein2}
\end{equation}
where a prime denotes $\partial/\partial r$.

The obtained power series form of solutions is \cite{BMcK}
\begin{equation}
2m = 4b^2 r^3 + {16 \over 5} b^3 r^5 + O(r^7),
\label{mso}
\end{equation}
\begin{equation}
w = 1 + br^2 + ( {4 \over 5}b^3 + {3 \over 10}b^2) r^4 + O(r^6),
\label{wso}
\end{equation}
where $b$ is a parameter, and some discrete set of values are
taken for $b$ in the range $ -0.706 < b < 0$. The solutions
(\ref{mso}) and (\ref{wso}) are regular in the whole range of
$r$, i.e., $0 \leq r < \infty $. The solutions can be
classified by the zeroes (n) of $w_n$, where $n$ can take all
the odd positive integers.

It is easy to show that the density $T_{00}$ is regular even at
$r = 0$, which is different from the Schwarzschild solution and
the Reissner-Norstrom solutions. The spacetimes are asymptotically
flat. The spacetimes have no event horizon at all. This fact
by no means violates Birkhoff theorem, because here the solutions
are not for the vacuum
Einstein equation.

By the equations (\ref{mso}) and (\ref{wso}), we obtain the
energy-momentum tensor
\begin{equation}
T^0_0 = -6b^2e^2{\kappa}^{-2} + O(r^2),
\label{em0}
\end{equation}
and
\begin{equation}
T^i_k = 2b^2e^2{\kappa}^{-2}{\delta}^i_k + O(r^2), \ \ \ for \ \ \ i,
k = 1,2,3,
\label{em1}
\end{equation}
for small $r$. Here we have already written the energy-momentum tensor
(\ref{em0}) and (\ref{em1}) in terms of the unscaled $r$. It
is worthy of noticing that this energy-momentum
tensor approaches that of perfect liquid with $p = {\epsilon}/3
= 2b^2e^2{\kappa}^{-2}$ in the FRW spacetime as $r$ approaches zero. Thus
the metric can also be set into the same as the metric of a FRW
spacetime near the spatial origin approximately. This is a very
important fact for our purposes.

To understand the global properties, we should give the Penrose
diagram of the BK spacetime. For this purpose, it is sufficient
to consider the two-dimensional metric,
\begin{equation}
ds^2 = -T^{-2} dt^2 + R^2 dr^2. \\
\label{metp}
\end{equation}
The null geodesic equation is
\begin{equation}
0 = g_{ab}k^a k^b = -T^{-2}{ \dot{t}}^2 + R^2{\dot{r}}^2.
\end{equation}
Thus, the radial null geodesic of BK spacetime satisfies
\begin{equation}
t = \pm r_* + constant,
\end{equation}
where
\begin{equation}
r_* = {\int}^r R(r^{\prime}) T(r^{\prime}) dr^{\prime}.
\end{equation}

We define the null coordinates $u$, $v$ by
\begin{equation}
u = t - r_*,
\label{tr1}
\end{equation}
and
\begin{equation}
v = t + r_*.
\label{tr2}
\end{equation}
In terms of the null coordinates, the metric (\ref{metp}) can
be written as
\begin{equation}
ds^2 = - T^{-2}(u,v) du dv.
\label{metn}
\end{equation}
Because both of the function $ T^{-2}(u,v)$ and the transformations
(\ref{tr1}) and (\ref{tr2}) are always regular, the metric
(\ref{metn}) is conformal to those of the Minkowski spacetime
suppressing the metric for the two-sphere. The Penrose diagram
is exactly the same as that of the Minkowski spacetime.

\vspace{5mm}
%%%%%%%%%%%%%%%%%%%%%%%%%%%%%%%%%%%%%%%%%%%%%%%%%%%%%%%%%%%%%%%%%%%%%%%%%%%%%%%
%%%%%%% Section 3. The Dynamics of Bubbles between BK and %%%%%%%%%%%%%%%%%%%%%
%%%%%%%                      YMC Spacetimes               %%%%%%%%%%%%%%%%%%%%%
%%%%%%%%%%%%%%%%%%%%%%%%%%%%%%%%%%%%%%%%%%%%%%%%%%%%%%%%%%%%%%%%%%%%%%%%%%%%%%%
\section{Dynamics of Bubbles between BK and YMC Spacetimes}

We shall discuss the dynamics of a bubble between BK and
YMC spacetimes and examine if a bubble can be created without
a singularity. This is intuitively
possible because the BK spacetimes, unlike the Schwarzschild
spacetime with spacelike singularity at $r=0$ in which the bubble
necessarily has an initial singularity \cite{FG,Guth}, is regular in the
whole spacetime even at $r=0$. However, we further ask what are the
properties of the bubble. Especially, can it expand large enough
to become a child universe? In this section we shall give a
quantitative discussion.

As is well known, a suitable tool for the research of the
evolution of a thin shell is what is called the
junction conditions on the shell which correspond to the Einstein
field equation together with conservation law of the
energy-momentum tensor. We shall summarize the results in the
subsection {\it a} \cite{BGG,BKT2}.

Before we can solve the junction condition equations, we should
know the form of energy-momentum tensor on the shell. Provided
that the shell is sufficiently thin, we can simply assume that it
has a $\delta$ function distribution across the shell. In the
subsection ${\it b}$, we shall derive the equation of motion
of the thin shell. Considering that the shell is made of the
Yang-Mills field, rather than scalar fields, the
properties of it should greatly differ from the case of a scalar
shell, we shall discuss it in detail in the subsection {\it c}
near the creation, and the subsection {\it d} for the final
stage, respectively.

\vspace{5mm}
\noindent
{\it a. The Einstein field equation on the thin shell and
Junction conditions}
\vspace{5mm}

To describe the behavior of the shell which divides the
spacetime into two regions $V^+$ and $V^-$, it is simplest
to introduce a Gaussian normal coordinate system in the
neighborhood of the shell. If we denote by $\Sigma$ the
(2+1)-dimensional spacetime hypersurface in which the shell
lies, we can introduce (2+1)-dimensional coordinates $x^i
\equiv (\tau, x^2, x^3) (i=1, 2, 3)$ on $\Sigma$. For the
timelike coordinate we use the
proper-time variable $\tau$ that would be measured by an
observer comoving with the shell. In the Gaussian normal
coordinate system, a geodesic in a neighborhood of $\Sigma$
which is orthogonal to $\Sigma$ is taken as the third spatial
coordinate denoted by $\eta$. Thus, the full set of
coordinates is given by $x^{\mu} \equiv (x^i,\eta)$. We can
take $x^2 = \theta$ and $x^3 = \phi$ in the case of
spherically symmetric shell. Then the metric can be written as
\begin{equation}
ds^2 = d{\eta}^2 + g_{ij}dx^i dx^j.
\label{me1}
\end{equation}
We introduce an unit vector field ${\xi}^{\mu}(x)$ which is
normal to $\eta$ = const. hypersurfaces and pointing
from the YMC to the BK spacetime. The induced metric on the
hypersurface $\Sigma$ can then be written as
\begin{equation}
h_{\mu \nu} = g_{\mu \nu} - {\xi}_{\mu} {\xi}_{\nu}.
\label{ind}
\end{equation}
In the Gaussian normal coordinates,
\begin{equation}
{\xi}^{\mu}(x) = {\xi}_{\mu}(x) = (0, 0, 0, 1)
\label{geo}
\end{equation}

The Gaussian normal coordinate system is suitable for the
Gauss-Codazzi formalism in the neighborhood of $\Sigma$ with the
coordinate $\eta$ orthogonal to the slices near $\Sigma$. The
extrinsic curvature is given by
\begin{equation}
K_{ij} = -{\Gamma}^{\eta}_{ij} = {1 \over 2}
{\partial}_{\eta}g_{ij}.
\label{ext}
\end{equation}

Assuming that the energy-momentum tensor $T_{\mu \nu}$ is of the
form
\begin{equation}
T_{\mu \nu} = S_{\mu \nu} {\delta}(\eta) + (regular \ \  terms),
\label{eng}
\end{equation}
where $S_{\mu \nu}$ is the surface stress-energy tensor, we can
rewrite the Einstein field equation as
\begin{equation}
[{K_j}^i] = - \kappa (S_j^i - {1 \over 2} {\delta}_j^i S),
\label{jun}
\end{equation}
\begin{equation}
{S_i}^j_{|j} = -[{T_i}^\eta],
\label{ei1}
\end{equation}
and
\begin{equation}
\{{K_j}^i\} {S_j}^i = [{T_\eta}^\eta],
\label{ei2}
\end{equation}
where $\{K_{ij}\} = {1 \over 2}( K_{ij}^+ + K_{ij}^-)$,
and $[\Psi] = {\Psi}^+ - {\Psi}^-$, hereafter for any quantity
$\Psi$.

By the symmetry analysis and the energy-momentum conservation
\cite{BGG}, the surface stress-energy tensor on the shell
should be
\begin{equation}
S^{\mu \nu} = {\sigma}(\tau) U^{\mu} U^{\nu} - {\zeta}(\tau)
(h^{\mu \nu} + U^{\mu} U^{\nu}),
\label{str}
\end{equation}
where $U^{\mu} = (1, 0, 0, 0)$ is the four-velocity of the shell. Here
$ {\sigma}$ is the surface energy density, and $\zeta$ is the surface
tensor of the shell. $\sigma$ and $\zeta$ should be constrained by the
Einstein field equation or conservation equation (see the equation
(\ref{eco}) below) and determined by the Yang-Mills equation.

\vspace{5mm}
\noindent
{\it b. The dynamics of the bubble between BK and YMC spacetimes}
\vspace{5mm}

For the sake of simplicity, we shall only consider a spherically
symmetric shell. In the coordinates of BK spacetime, the metric
is described by
\begin{equation}
ds^2 = - T^{-2} dt^2 + R^2 dq^2 + r^2(q, t) d{\Omega}^2_2.
\label{wal}
\end{equation}
The coordinate $q$ is defined so that it originates from the
center of the bubble to the outside. The induced metric on
the shell is determined by
\begin{equation}
ds^2|_{\Sigma} = - d{\tau}^2 + {\rho}^2(\tau) d{\Omega}^2_2,
\label{fbm}
\end{equation}
where $\tau$ is the proper time of the shell.

Inside the bubble is the domain of the YMC (FRW) spacetime with
the metric
\begin{equation}
ds^2 = - dt^2 + a^2(t)( d{\chi}^2 + d{\Omega}^2).
\label{frwp}
\end{equation}
The matter source is the Yang-Mills field in the YMC.

To the observer in this coordinate system, the shell corresponds
to $\rho = a \cdot f(\chi)$, $U^0 = U_0 = 1$, and $U^1 = U_1 = 0$.
The explicit form of the extrinsic curvature of the shell in
this case was already given in Ref. \cite{AKMS} and \cite{BKT2}. We
just take advantage of results there;
\begin{equation}
{K_2}^2 = - {\Sigma \over \rho}( {\dot{\rho}}^2 + 1 - {\kappa
  \over 3} \epsilon {\rho}^2)^{1/2},
\label{e22}
\end{equation}
where $\Sigma = 1$ for the case of the increasing shell radius
and $\Sigma =-1$ for the case of the decreasing shell radius
in the direction of the outer normal. Here, by the results in
the section 2, $\epsilon = {3C^2 \over 2e^2 a^4}$.

For the spherically symmetric bubble, the equation (\ref{ei1})
will give a relation
\begin{equation}
\dot{\sigma} = - 2 ( \sigma - \zeta ) {\dot{\rho} \over \rho}
-  g\rho \dot{\sigma},
\label{eco}
\end{equation}
where $g = {e \over \sqrt{\kappa}}$ and the overdot denotes a
derivative with respect to $\tau$.
Here we used the equations (\ref{tet}) and (\ref{tte}) in the
Appendix, where the form of $[{T_k}^\eta]$ has been
discussed for general spherically symmetric Yang-Mills shell
configurations.

For the shell, by the equations (\ref{mso}) and (\ref{wso})
and the relation $\rho = r$ on the shell, the 2-2 component of
the extrinsic curvature is obtained as
\begin{equation}
{K_2}^2 = - {\Sigma \over \rho} ( {\dot{\rho}}^2 +
1 - {2 m(\rho) \over \rho})^{1/2}.
\label{b22}
\end{equation}
in the coordinates of the BK spacetime.

Since in the spherically symmetric case $A^{22} = A^{33}$ for
any second-rank tensor $A^{\mu \nu}$, we need only consider the
2-2 component of the junction condition equation (\ref{jun})
\begin{equation}
- {{\Sigma}_{out} \over \rho} ({\dot{\rho}}^2 + 1 -
{2 m(\rho) \over \rho})^{1/2} +
{{\Sigma}_{in} \over \rho} ( {\dot{\rho}}^2 + 1 - {\kappa \epsilon
  \over  3} {\rho}^2)^{1/2} = {\kappa \over 2}\sigma.
\label{2-2}
\end{equation}

It is easy to show that the independent components of the
Einstein field equation and the conservation equation are the
two equations (\ref{eco}) and (\ref{2-2}) \cite{BKT2}. Now we
engage ourself in solving these equations.

We should have an initial condition for the shell. In the
neighborhood of the origin of the BK spacetime, the curvature
approaches that of YMC ( FRW ) spacetime (see Fig. 1). If we
suitably choose the initial scale factor $a_0$ and the
``energy'' $C^2$ of the Yang-Mills field in YMC spacetime so that
\begin{equation}
a_0 = ({{\kappa}^2C^2 \over 4 e^4b^2})^{1/4},
\label{ini}
\end{equation}
we can get a YMC bubble smoothly created from the BK spacetime,
because of the continuity of curvatures for the two spacetimes
on the shell. The meaning of it is that, for a fixed BK spacetime
namely for a given $b$, a small part of the YMC spacetime near the
origin satisfying the condition (\ref{ini}) may smoothly emerge out of
the BK spacetime and develop into a bubble. For the later development
of the bubble, we should solve the equations (\ref{eco}) and
(\ref{2-2}) with the initial condition (\ref{ini}).

However we still have too many unknown functions $\sigma$,
$\zeta$ and $\rho$, while we have only two independent
equations. We need an equation of state of the
shell from the Yang-Mills field equation. However
for thin shells, the simplest class of the equation of
state may be of the form $\zeta = (1 - \alpha ) \sigma $.
Under this condition, we can easily solve Eq. (\ref{eco})
\begin{equation}
\sigma = C^{\prime}({\rho \over 1 + g\rho})^{-2 \alpha},
\label{sig}
\end{equation}
and
\begin{equation}
\zeta = C^{\prime}(1 - \alpha) ({\rho \over 1 + g\rho})^{-2 \alpha}.
\label{zet}
\end{equation}
Here $C^{\prime}$ is a constant with unit $[L]^{2\alpha -3}$. For the
case without initial singularity,
we should have $\alpha < 0$, because at initial point the YMC bubble
is a part of the BK spacetime near $r = 0$ and there is no shell at
all there.

Substituting the surface energy $\sigma$ given by Eq.(\ref{sig})
into Eq. (\ref{2-2}), we obtain
\begin{equation}
- {\Sigma}_{out} ({\dot{\rho}}^2 + 1 - {2m(\rho) \over \rho})^{1/2} +
{\Sigma}_{in} ( {\dot{\rho}}^2 + 1 - {\kappa \epsilon \over 3}
{\rho}^2)^{1/2} = {\kappa C^{\prime} \over 2} {{\rho}^{1 - 2 \alpha}
\over (1 + g\rho)^{-2\alpha}}.
\label{2,2}
\end{equation}
Twice squaring the equation (\ref{2,2}), we get the equation of
motion for the shell,
\begin{equation}
{\dot{\rho}}^2 + V(\rho) = 0,
\label{mot}
\end{equation}
where
\[
V(\rho) = {(1+g\rho)^{-4\alpha} \over (\kappa C^{\prime})^2
          {\rho}^{2-4 \alpha}}[4 (1 - {2m(\rho) \over \rho}) (1 -
          {\kappa \epsilon \over 3}{\rho}^2 ) - ( 2 - {2m(\rho) \over \rho} -
          {\kappa \epsilon \over 3} {\rho}^2
\]
\begin{equation}
           -({\kappa C^{\prime} \over 2})^2 {{\rho}^{2 - 4 \alpha} \over
          (1+g\rho)^{-4\alpha}})^2].
\label{POT}
\end{equation}
Then the equation of motion for the shell is identical to that of
a classical particle moving in one-dimension under the influence
of the potential (\ref{POT}), with the energy zero.

For the convenience of the later discussions, we write the
potential (\ref{POT}) as
\begin{equation}
V(\rho) = 1 - {2m(\rho) \over \rho} - {1 \over 4 B}(A - B)^2,
\label{POTp}
\end{equation}
where
\begin{equation}
A = {2m(\rho) \over \rho} - {\kappa \epsilon \over 3} {\rho}^2,
\end{equation}
and
\begin{equation}
B = ({\kappa C^{\prime} \over 2})^2 {{\rho}^{2 - 4 \alpha} \over
(1+g\rho)^{-4\alpha}}.
\end{equation}

\vspace{5mm}
\noindent
{\it c. The properties near the creation }
\vspace{5mm}

We are especially interested in the case without an initial
singularity. In the language of the potential (\ref{POT}), a
good behavior of it is necessary at the point near $\rho = 0$.
However, it is perhaps impossible to analytically solve the
equation of motion with the potential (\ref{POTp}). This is
not only because of the complexity of the potential but also
because of the proper time dependence of $\epsilon (\tau)$ of
the potential. We shall give a numerical analysis in the
forthcoming paper. Here let us analyze some special cases. We are
especially interested in the case of the very small bubble
because we want to know whether the initial singularity can be
avoided or not. If we only consider a bubble developing in a
short duration comparing with the whole life of the YMC
universe, we can approximately take $\epsilon$ as a constant.
For our purpose it is sufficient to consider a special case
$\alpha = - 1/2$. The potential (\ref{POTp}) is equal to
\begin{equation}
V(\rho) = 1 - {2m(\rho) \over \rho} - {{\beta}^2 \over 4},
\label{POT1}
\end{equation}
where ${\beta} \equiv {A \over B^{1/2}}$ is a constant because as
$\rho \rightarrow 0$, $A^2, B \propto {\rho}^4$, and $\epsilon
\rightarrow {3C^2 \over 2e^2a_0^4}$. In the
equation of motion (\ref{2,2}), we have two independent arbitrary
constants $C^{\prime}$ and $b$ (see equation (\ref{mso})). The
constant $\beta$ can be written in terms of $C^{\prime}$ and $b$,
\begin{equation}
\beta = {8e^2 b^2 - { 2/3{\kappa}^2\epsilon } \over
{\kappa}^2 C^{\prime}},
\end{equation}
with $\alpha = - 1/2 $. If the shell satisfies the condition
\begin{equation}
{\beta}^2 \geq 4,
\label{con}
\end{equation}
remarkably, the potential is finite at $\rho = 0 $, and a
shell is always classically possible. This finiteness is
what is necessary. Otherwise, there is an inconsistency
if there is no initial singularity.

In fact the value of the $\beta $ can be determined uniquely by the
initial velocity of the shell. For a shell with a zero velocity (or
infinitesimal velocity), we have $\beta = 4$ (or $\beta \rightarrow
4$). A typical potential is shown in Fig. 2 corresponding to a BK
spacetime of $b = - 0.45371627277$, and ${\beta}^2 = 4$.
\vspace{5mm}

\noindent
{\it d. The properties of the final stage}
\vspace{5mm}

In general case, of course, $\epsilon$ is time dependent, because it
is the energy density in the YMC (FRW) spacetime. Although we cannot
solve the equation of motion (\ref{mot}) analytically, we can show an
important property for the trajectory of shell. Let us consider the
potential (\ref{POTp}). By the properties of the BK solution, one can
find $2m^{\prime}(\rho) < {6m(\rho) \over \rho}$. This fact means that
${m(\rho) \over {\rho}^3}$ monotonically decreases with $\rho$
increasing. Thus the last term in the potential (\ref{POTp})
monotonically decreases with $\rho$ increasing. As we known, when
$\rho \rightarrow 0$, the last term approaches a constant. Since the
constant is smaller or equal to $1$ and the second term is definitely
negative, the potential is definitely negative for all $\rho > 0$.
This means that the velocity of the shell cannot change the sign in
the evolution. The shell always increases since the initially it
increases. Thus the bubble cannot be the type of $k = 1$, since the
bubble would collapse finally. This fact depends on the special type of
shell with $\alpha = -1/2$. Though both of $k = 0$ and $k
= -1$ types of bubbles are possible (be determined by the initial spatial
curvature), here we only discuss the case of $k = -1 $ in detail. The
discussion for the case of $k = 0$ is similar.

Finally $\rho$ will become very large, and the potential
(\ref{POTp}) can be written as
\begin{equation}
V \simeq - {2m(\rho) \over \rho} - {(\kappa C^{\prime})2 \over 16}
{{\rho}^4 \over (1+g\rho)^2},
\label{lp}
\end{equation}
approximately. Thus with the increasing of $\rho$, $V$ will
decrease to negative infinity, which makes $\dot{\rho} \rightarrow \infty$. In
terms of $\dot{\rho}$, we can describe the velocity of the shell as
\begin{equation}
({d\rho \over {dt}})_{BK} = {\dot{\rho} \over  T^2\sqrt{1 +
R^2\dot{\rho}^2 }},
\end{equation}
with respect to the observer inside the BK spacetime, and
\begin{equation}
({d\chi \over {d\lambda}})_{YMC} = {\dot{\rho} \over \sqrt{1 +
\dot{\rho}^2 }},
\end{equation}
with respect to the observer inside the YMC spacetime. The
trajectory of the shell approaches null with respect to both of the
observers, as the shell becomes infinity.

We can understand the fact better by the Penrose diagrams.
Since the scale factor $a^2(\lambda)$ is a conformal factor in
the metric (\ref{frw}), which will not show up in the Penrose
diagram, $\dot{\rho}$ is just equivalent to $\dot{\chi}$ in terms
of the conformal metric $ ds^2 = -d{\lambda}^2 + d{\chi}^2 +
f^2(\chi)d{\Omega}^2 $.
However, by the equation (\ref{lp}), $\rho$ always increases
at the later stage. Thus the shell will always increase in
the $\chi$ parameter space of spatial hypersurface $H^3$. The
trajectory of the shell approaches null and finally intersects $I^+$.
The situation is shown in the Fig. 3.

For an observer in the region of BK spacetime, because $\rho$
corresponds the usual radial coordinate, the shell always
increases too (see Fig. 3).

We also give a combined figure as Fig. 4.

The situation described above is totally different from the case of a
de Sitter bubble in the Schwarzschild spacetime \cite{BGG,FG,Guth}.

\vspace{5mm}
%%%%%%%%%%%%%%%%%%%%%%%%%%%%%%%%%%%%%%%%%%%%%%%%%%%%%%%%%%%%%%%%%%%%%%%%%%%%%%%
%%%%%%%% Section 4. Relations to Singularity Theorems        %%%%%%%%%%%%%%%%%%
%%%%%%%%%%%%%%%%%%%%%%%%%%%%%%%%%%%%%%%%%%%%%%%%%%%%%%%%%%%%%%%%%%%%%%%%%%%%%%%
\section{Relations to Singularity Theorems}
It is well known that there are a series of theorems which state
the existence of a singularity in a spacetime under some physically
reasonable conditions \cite{P,HP,HE,W}. Why we can avoid an initial
singularity for a YMC bubble inside a BK spacetime? In the
introduction we already gave an intuitive discussion about this. Now
we shall give some further discussions relevant to the singularity
theorems.

At first, we should examine the energy conditions because our
spacetime contains a special form of matter -- shell, although
the regions of the BK and YMC spacetimes satisfy the strong energy
condition. Let us consider a general shell with the surface
stress-energy tensor (\ref{str}). Since
\begin{equation}
T^{\mu \nu} k_{\mu} k_{\nu} = \sigma (k_{\tau})^2 \delta (\eta) \geq 0,
\end{equation}
for any null vector $k^{\mu}$, the very weak energy condition \cite{FG}
is satisfied. To check the strong energy condition, we may consider
the quantity,
\begin{equation}
T^{\mu \nu} U_{\mu} U_{\nu} + {1 \over 2} T = {1 \over 2} (\sigma - 2 \zeta)
\delta (\eta),
\end{equation}
for the 4-velocity $U_{\mu}$ of the shell. We find the strong
energy condition fails on the shell for a repulsive shell,
$\sigma - 2 \zeta \leq 0$, in which we are interested.

We then find no contradiction between our result and the singularity
theorem of Hawking and Penrose \cite{HE}, since the theorem needs
the strong energy condition. However, the singularity theorem of
Penrose needs only very weak energy condition which is satisfied
on the shell. We should answer why we can avoid this theorem
too. The Penrose theorem tells us that

\vspace{5mm}
\noindent
spacetime ($ \cal{ M }$, ${\bf g}$ ) cannot be past-null
geodesically complete if

\noindent
(1) $R^{\mu\nu}k_{\mu}k_{\nu} \ge 0 $ for all null $k_{\mu}$,

\noindent
(2) there is  a noncompact Cauchy surface in $\cal{M}$, and

\noindent
(3) there is an anti-trapped surface in $\cal{M}$.

\vspace{5mm}
\noindent
Here we have expressed the original theorem slightly differently.
Namely we replace the trapped surface by the anti-trapped surface,
for the purpose of the discussion of an initial singularity.

In the paper of Farhi and Guth \cite{FG}, the authors argued
that an initial singularity necessarily exists because the
de Sitter bubble inside the Schwarzschild spacetime has
anti-trapped surfaces. Although there exists an initial
singularity in that case, we should notice at least two points.
First, we cannot apply the Penrose theorem only to the de Sitter
bubble region because there is no well-defined Cauchy surface.
Nor can we use its modification in which the second condition
is replaced by the stable causality and the generic condition
\cite{GH}, because the de Sitter spacetime does not satisfy
the latter one. To define a Cauchy surface, we need to consider
the combination of the bubble and parent spacetimes. For example,
a Cauchy surface is shown in the Fig. 5 in the case of a de
Sitter-Schwarzschild bubble. Second, the bubble itself is not
an inextendible spacetime. However a total combination of bubble,
parent spacetimes and the shell between them can be considered
as an inextendible spacetime to which the singularity theorem
applies. Thus we can apply the Penrose theorem to this combination
since it is inextendible and permitted to define a Cauchy surface.

In the proof of his theorem, Penrose defined a closed
trapped surface \cite{P}. A significant feature of a trapped
surface $T$ arises from the fact that the null geodesics
meeting it orthogonally are the generators of horismos $E^+(T)$
(or $E^-(T)$) \cite{HP}. On the surface, these null geodesics start
out by converging. If the weak energy condition and null
completeness are assumed, the Raychaudhuri equation for the null
geodesic tells us that they must continue to converge until
encountering focal points finally. The geodesic segments
joining $T$ to the focal point must sweep out a compact set. By
this means $T$ is called future ( or past ) trapped set \cite{HP}.

However, the relation between trapped surface and trapped set
becomes subtle if a shell exists. It is instructive to analyze
again the reason why the de Sitter-Schwarzschild bubble has
an initial singularity. A null geodesic starting from a trapped
surface (if exists in one region of the spacetime) will penetrate
the shell and enter into another region generally. Then we can
determine whether this trapped surface is a future trapped set
or not by looking at the whole spacetime, but cannot determine
by only looking at the bubble or the parent spacetimes separately.
For a de Sitter bubble inside the Schwarzschild spacetime, we can
easily find an anti-trapped surface. One of the anti-trapped
surfaces is shown in Fig. 5. The outer directed $E^-(T)$ passes
through the shell and go into the region of Schwarzschild
spacetime. However, this $E^-(T)$ will also converge in the
region of white hole and form a focal point, if completeness of
null geodesic is assumed. Thus this anti-trapped surface is a
past trapped set. Similarly to the proof of the Penrose theorem,
this completeness would make contradiction between the
compactness of horismos $E^-$ and the non-compactness of the
Cauchy surface, and then there must be a singularity -- null
geodesics should not be complete. Here we should notice that
for an initial singularity we should have an anti-trapped
surface, and for a final or collapsing singularity, we should
have a trapped surface.

A YMC (FRW) bubble created inside of the BK spacetime looks
similar to that of the de Sitter case. However, a fundamental
difference is that the anti-trapped surface region is in a different
area in the corresponding Penrose diagrams. In a FRW universe
(e.g. $k = - 1$), there always exists an anti-trapped surface if
\begin{equation}
{d \over  dt}(a^2(t) sinh^2(\chi)) > 0
\label{tra}
\end{equation}
holds for both families of inward and outward null geodesics
\cite{HE}.  For the convenience of our discussions, we should
know the region in the Penrose diagram where anti-trapped
surfaces exist. To a spacetime which is invariant under action of
translations, e.g., a whole FRW universe, the inequation (\ref{tra})
does not serve to give a specific spatial region of anti-trapped
surfaces. However, since we are discussing a bubble which develops
from an infinitesimal bubble at the origin of the BK spacetime, we can
specify an anti-trapped region with respect to a point which
corresponds to the origin of the BK spacetime. In terms of conformal
time $\lambda$, we can write the equation (\ref{tra}) as
\begin{equation}
{1 \over tanh({\lambda}_0)} > \pm{1 \over tanh({\chi}_0)},
\end{equation}
where the subscript $0$ corresponds to a two-sphere. Here we have used
the condition that the radius of the bubble ${\chi}_0 = 0$ at the time
${\lambda}_0 = 0$. This inequality is equivalent to
\begin{equation}
{\lambda}_0 < {\chi}_0,
\end{equation}
where ${\lambda}_0 = {\chi}_0$ corresponds to the event
horizon with respect to the observer at the canter of the bubble
(Although the apparent horizon is typically a null
or spacelike surface which lies inside or coincides with the
event horizon assuming cosmic censorship, we cannot use this result
directly because the cosmic censorship does not apply for the child
universe. However, by the calculation above, we showed that this
result is also correct for our case even without
assuming cosmic censorship). Thus an anti-trapped surface
always exists beyond the event horizon (to the observer at origin)
of the YMC (FRW) spacetime.

To summarize, as shown in the Penrose diagrams (see Fig. 5),
anti-trapped surface region is III and trapped surface region
is IV for the de Sitter spacetime, while there is no trapped surface
and the anti-trapped region is II for the open FRW spacetime (see
Fig. 3).

It is easy to see that any de Sitter bubble with a non-spacelike shell
can enter the region with an anti-trapped surface. Then there must be
an initial singularity, if other conditions of the Penrose theorem
are also satisfied.

In the case of the YMC bubble in a BK spacetime, a bubble can be
parameterized by $sinh( \chi)$ in a spacetime with metric  $ ds^2 =
-d{\lambda}^2 + d{\chi}^2 + sinh^2 (\chi)d{\Omega}^2 $ which is conformal
to YMC spacetime. The developing of the bubble corresponds the change
of radius ${\chi}_S$ of the shell. A bubble may or may not contain an
anti-trapped surface. If the motion of a shell is non-spacelike, a
bubble which develops from the origin of the BK spacetime can never
enter the region with an anti-trapped surface. In fact, any past
directed null geodesic starting from point inside the bubble will go
to null infinity $I^-$ in the BK spacetime after penetrating the
shell. Thus there exists no two-surface which is a past trapped set.
This is the reason of the absence of initial singularity.

Because there is no anti-trapped surface, there is also no
contradiction to the singularity theorems with generalized
strong energy conditions \cite{T,B}.

\vspace{5mm}
%%%%%%%%%%%%%%%%%%%%%%%%%%%%%%%%%%%%%%%%%%%%%%%%%%%%%%%%%%%%%%%%%%%%%%%%%%%%%%%
%%%%%%%%%%% section 5. Conclusion and Discussions  %%%%%%%%%%%%%%%%%%%%%%%%%%%
%%%%%%%%%%%%%%%%%%%%%%%%%%%%%%%%%%%%%%%%%%%%%%%%%%%%%%%%%%%%%%%%%%%%%%%%%%%%%%%
\section{Conclusions and Discussions }
In this paper, we have discussed the properties of YMC-BK
bubble and given an example to show how a YMC bubble emerges
from a BK spacetime without an initial singularity. By
contrasting to the case of a de Sitter bubble created from
a Schwarzschild spacetime, we have also given an intuitive
picture as depicted in Fig. 6 which shows snap shots of the
four characteristic epochs. Comparing to the de-Sitter-Schwarzschild
bubble, a fundamental difference is that the bubble can never be
separated from the parent spacetime.

Coincidently our results can just describe the instability of the
outside BK spacetime.
At early stage of the bubble, our results qualitatively coincide to
that of ``mini star'' case which was discussed by Zhou and Straumann
numerically \cite{ZS}. However, their discussion could not
be extended to the later stage. The reason is that, with the
evolution, the Yang-Mills field configuration of the ``mini
star'' die away more and more rapidly as increasing of $r$.
While the numerical method is not suitable, thin shell method
is still effective to such a case. The thin shell formalism could
give an analytical discussion from which we could give the global
properties of the YMC-BK bubble which corresponds to a ``mini
star'' spacetime at early stage.

There are two kinds of YMC bubbles which can develop inside the
BK spacetime. The shells are exact same even when the inside spacetimes
are different. The type of the bubble is further constrained by the
constant $C^2$. If $0 \leq C^2 \leq 1$, the bubble should uniquely
be a part of the YMC spacetime with $k=0$. No $k = -1 $ type of FRW
solution if $0 \leq C^2 \leq 1$ \cite{GV1}. If $C^2 >1$, both $k=0$
and $k = -1$ types of bubble are possible.

Because the shells expand not only toward the YMC but also toward
the BK spacetimes, comparing to a de Sitter bubble in a
Schwarzschild spacetime, the developing of a YMC bubble in a
BK spacetime is less ambiguous because there is not any paradox
here. The observer on the side of YMC spacetime near the shell
attributes the expansion of the shell to the expansion of the
YMC spacetime and high pressure comparing to the outside. The
outer observer near the shell attributes this to that the pressure is
lower than the YMC bubble.

%%%%%%%%%%%%%%%%%%%%%%%%%%%%%%%%%%%%%%%%%%%%%%%%%%%%%%%%%%%%%%%%%%%%%%%%%%%%%%%
%%%%%%%%%  Appendix. The Form of [{T_i}^{eta}] between Spherical Shells %%%%%%%
%%%%%%%%%%%%%%%%%%%%%%%%%%%%%%%%%%%%%%%%%%%%%%%%%%%%%%%%%%%%%%%%%%%%%%%%%%%%%%%
\vspace{5mm}
\renewcommand{\theequation}{A\arabic{equation}}
\begin{section}*{Appendix:
The Form of $[{T_i}^{\eta}]$ on Spherical Yang-Mills field Shells}

Because of the existence of the global time coordinate both for inside
and outside regions, it is easer to calculate the term $[{T_i}^{\eta}]$
in the case of a shell of a finite thickness than the one in the
previous infinite thin shell case. We shall first consider the shell
configuration exactly. Then by taking a limit of
zero thickness, we can get $[{T_i}^{\eta}]$ for the thin
shell. In this Appendix, we shall give the general form of
$[{T_i}^{\eta}]$ for a spherically symmetric, Yang-Mills field
shell.

For a developing shell, the Yang-Mills field and spacetime are
time dependent. The metric for the spacetime with such a shell
can be written as
\begin{equation}
ds^2 = -e^{2\psi} dt^2 + e^{2\lambda} dr^2 + r^2d{\Omega}^2,
\end{equation}
where $\psi$ and $\lambda$ are the functions of $r$ and $t$.
The Yang-Mills connection 1-form is assumed to be \cite{BMcK}
\begin{equation}
e{\bf A} = u{\sigma}_3 dt + w{\sigma}_1 d{\theta} +
(cot {\theta} {\sigma}_3 + w{\sigma}_2) sin {\theta} d{\phi}, \\
\end{equation}
where $u$ and $w$ depend only on $r$ and $t$. By the scale
transformation $r \rightarrow {\sqrt{\kappa} \over e}r$, the Einstein
and Yang-Mills equations can be transformed into $e$  and $\kappa$
independent for convenience of calculations.

The Yang-Mills field components are
\begin{equation}
B_T^2 = {e^{-2\lambda} \over r^2}(w^{\prime})^2,
 \ \ \ \
B_L^2 = {(1-w^2)^2 \over r^4},
\end{equation}
and
\begin{equation}
E_K^2 = {e^{-2\psi} \over r^2}\dot{w}^2,
 \ \ \ \
E_T^2 = e^{-2(\psi + \lambda)} {u^{\prime}}^2,
 \ \ \ \
E_L^2 = {e^{-2\psi} \over r^2}u^2 w^2.
\end{equation}
Here the prime denotes partial derivative with respect to $r$.
In this Appendix, we always denotes $\partial/\partial t$ by an
overdot, which is different in previous sections.

The non-vanishing energy-momentum tensor components for the
Yang-Mills field are
\begin{equation}
{T_0}^0 = - [B_T^2 + {1 \over 2}B_L^2 + E_K^2 + {1 \over 2} E_T^2 +E_L^2],
\label{tt1}
\end{equation}
\begin{equation}
{T_1}^1 = B_T^2 - {1 \over 2}B_L^2 + E_K^2 - {1 \over 2} E_T^2 +E_L^2,
\label{tt2}
\end{equation}
and
\begin{equation}
{T_0}^1 = {2\dot{w}w^{\prime} \over r^2} e^{-2\lambda}.
\label{tt3}
\end{equation}

Zhou and Straumann found the numerical solution for this system
\cite{ZS}. In terms of $m$ introduced by the relation
\begin{equation}
e^{-2\lambda} = 1 - {2m(r,t) \over r},
\end{equation}
similarly to the case in BK solution, these solution can be written
as
\begin{equation}
m(r,t) = 2b(t)^2r^3 + O(r^5),
\label{mt}
\end{equation}
and
\begin{equation}
w(r,t) = 1 + b(t)r^2 + O(r^4),
\label{wt}
\end{equation}
near the spatial origin. It is worthy of noticing that this solution
becomes the BK solution if $b(t) = b$ is a constant.

By the equations (\ref{mt}) and (\ref{wt}), the energy-momentum
tensor components (\ref{tt1}), (\ref{tt2}) and (\ref{tt3}) can
be written as
\begin{equation}
{T_0}^0 = -6b(t)^2 + O(r^2),
\end{equation}
\begin{equation}
{T_k}^i = 2b(t)^2 {\delta}_k^i + O(r^2),
\end{equation}
and
\begin{equation}
{T_0}^1 = 2 (b(t)^2)^.r + O(r^3),
\end{equation}
near the origin. It is easy to find that the spacetime is still
the YMC ( FRW ) type at the origin.

Because we are discussing a thick shell configuration
which connects the YMC spacetime near the origin to the outside
BK spacetime, we have
\begin{equation}
[{T_0}^1] \equiv {T_0}^1(r \rightarrow r_{BK}) -
{T_0}^1(r \rightarrow r_{YMC}) = - 2((b(t)^2)^. r_{YMC},
\end{equation}
because ${T_0}^1(r \rightarrow r_{BK}) = 0$. Here $r_{YMC}$ is
the corresponding radius of the YMC spacetimes
where the shell configuration connects to.

Now let us return to the thin shell case. By the definition
\cite{BKT1}
\begin{equation}
{S_k}^i = \lim_{\delta \rightarrow 0}\int d{\eta} {T_k}^i =
\sigma \delta(\eta) {\delta}_k^i,
\end{equation}
we have $\sigma = 2 b^2(t) - (2 b^2 + O(r^2))_{BK}$, since the
region $r < r_{YMC}$ can be regarded as the YMC spacetime region. Here
$\delta = 2\epsilon$ is the thickness of the shell. Considering the
time independence of the terms $(2 b^2 + O(r^2))_{BK}$ in $\sigma$, we
have
\begin{equation}
[{T_0}^1] = - \dot{\sigma} r.
\label{t01}
\end{equation}

Taking the approximation that the thickness of the shell approaches
zero and denoting in terms of the proper time of the shell, we finally
obtain
\begin{equation}
[{T_\theta}^{\eta}]=[{T_\phi}^{\eta}] =0,
\label{tet}
\end{equation}
and
\begin{equation}
[{T_\tau}^{\eta}] = - {e \over \sqrt{\kappa}} \rho {d\sigma \over d\tau},
\label{tte}
\end{equation}
where $\eta$ is the Cauchy normal coordinate. Notice that we have
already written the equation (\ref{tte}) in a form including the
constants $e$ and $\kappa$ by the transformation $r \rightarrow {e
\over \sqrt{\kappa}}r$.
\end{section}
\vspace{5mm}
%%%%%%%%%%%%%%%%%%%%%%%%%%%%%%%%%%%%%%%%%%%%%%%%%%%%%%%%%%%%%%%%%%%%%%%%%%%%%%%
%%%%%%%%%%%%%%%%%%%%%%%% acknowledgments %%%%%%%%%%%%%%%%%%%%%%%%%%%%%%%%%%%%%%
%%%%%%%%%%%%%%%%%%%%%%%%%%%%%%%%%%%%%%%%%%%%%%%%%%%%%%%%%%%%%%%%%%%%%%%%%%%%%%%

\vspace{5mm}
{\large{\bf Acknowledgments}}
\par
The author would like to express his sincere thanks to Professor
A. Hosoya for some helpful discussions and careful reading of the
manuscript. He is also grateful to Prof. K. Maeda and Drs. T. Koike,
T. Mishima, N. Sakai, T. Torii and K. Watanabe for useful discussions.
This work was supported in part by the Japanese Government.
\eject

%%%%%%%%%%%%%%%%%%%%%%% References %%%%%%%%%%%%%%%%%%%%%%%%%%%%%%%%%%%%%%%%%%%%

\eject
\noindent
{\large{\bf Figures Captions }}
\vspace{3mm}

\noindent
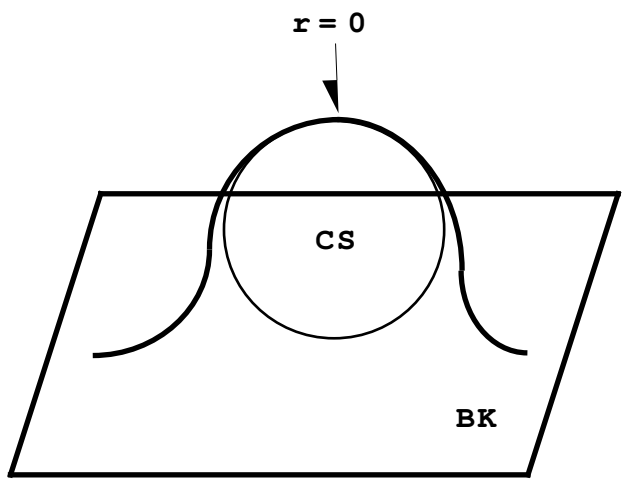
Fig. 1 \ \ \ The BK spacetime is equivalent to a YMC
spacetime near the origin.
\vspace{2mm}

\noindent
Fig. 2 \ \ \ The potential of the shell for the case
$b = - 0.45371627277$, and ${\beta}^2 = 4$.
\vspace{2mm}

\noindent
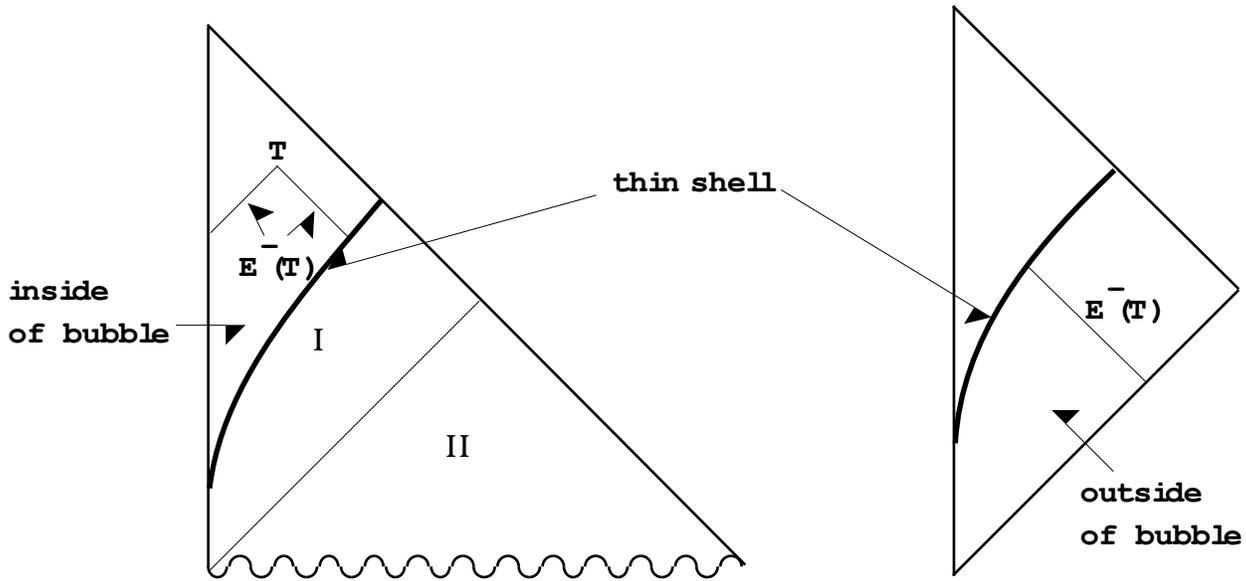
Fig. 3 \ \ \ The Penrose diagrams for the case of a YMC-BK bubble.
The one for the YMC (k = -1) spacetime is the same to those of FRW spacetime
in the radiation dominant epoch. The one for the BK spacetime is the same
to those of the Minkowski spacetime. The bubble cannot enter into the
region II, including an anti-trapped surface, since the motion of the
shell is non-spacelike. The solid line shows the trajectory of the shell.
\vspace{2mm}

\noindent
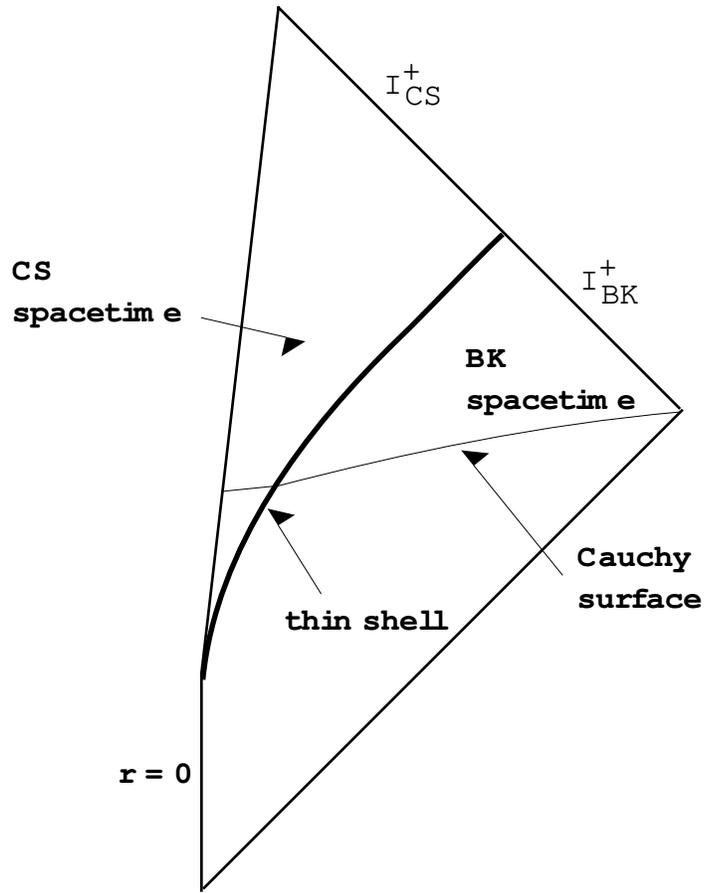
Fig. 4 \ \ \ The Penrose diagram with the inside and outside of
the bubble sewn together. The trajectory of the shell approaches
lightlike as the bubble expands to $I^+$.
\vspace{2mm}

\noindent
Fig. 5 \ \ \ The Penrose diagrams for the case of a
de Sitter-Schwarzschild bubble. There are anti-trapped
surfaces in the region III. $T$ is one of anti-trapped
surfaces whose past horismos $E^-(T)$ is compact if the
null completeness were supposed. A Cauchy surface is
also shown.
\vspace{2mm}

\noindent
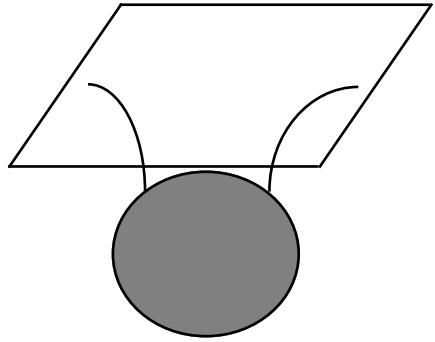
Fig. 6 \ \ \ Four characteristic times for the evolution of the
bubble. The picture {\it a} is a BK spacetime.


\begin{thebibliography}{100}
\bibitem{In}
     Here we should notice that such an initial singularity has
     no direct relation to
     the energy-momentum tensor singularity on the shell. Such
     an initial singularity is defined as that there exists a
     passed directed non-spacelike geodesic which has no endpoint
     or has a finite affine parameter length. In fact, we can
     find some spacetime which has infinite affine parameter
     length but has some form of singularity in $T_{\mu\nu}$.
     See, for example, the references \cite{HE,W}.
\end{thebibliography}
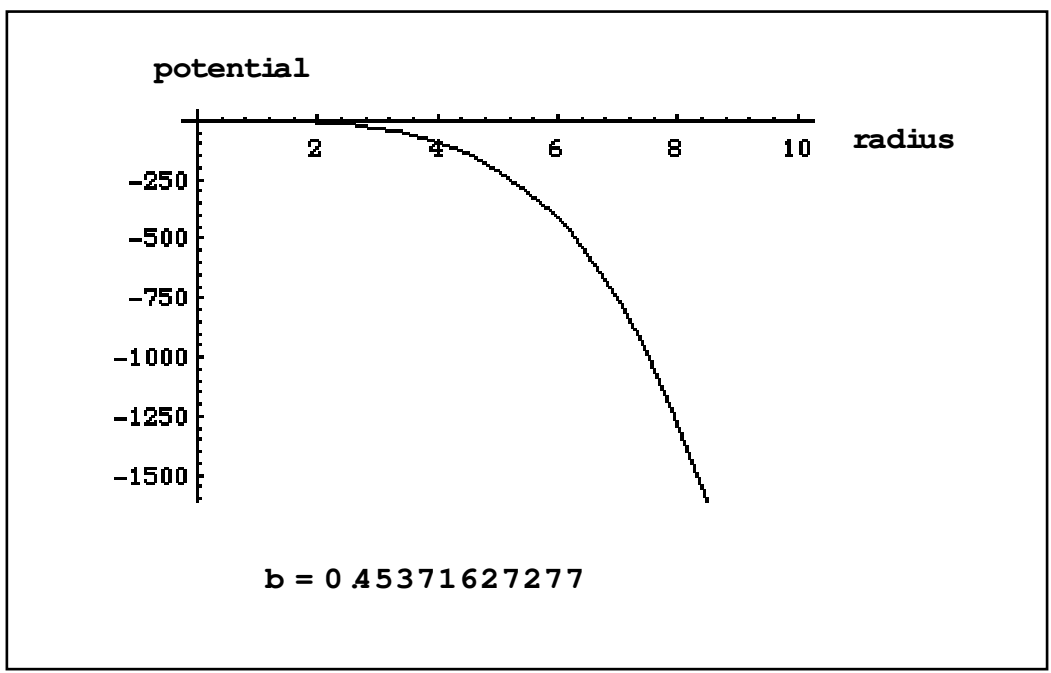
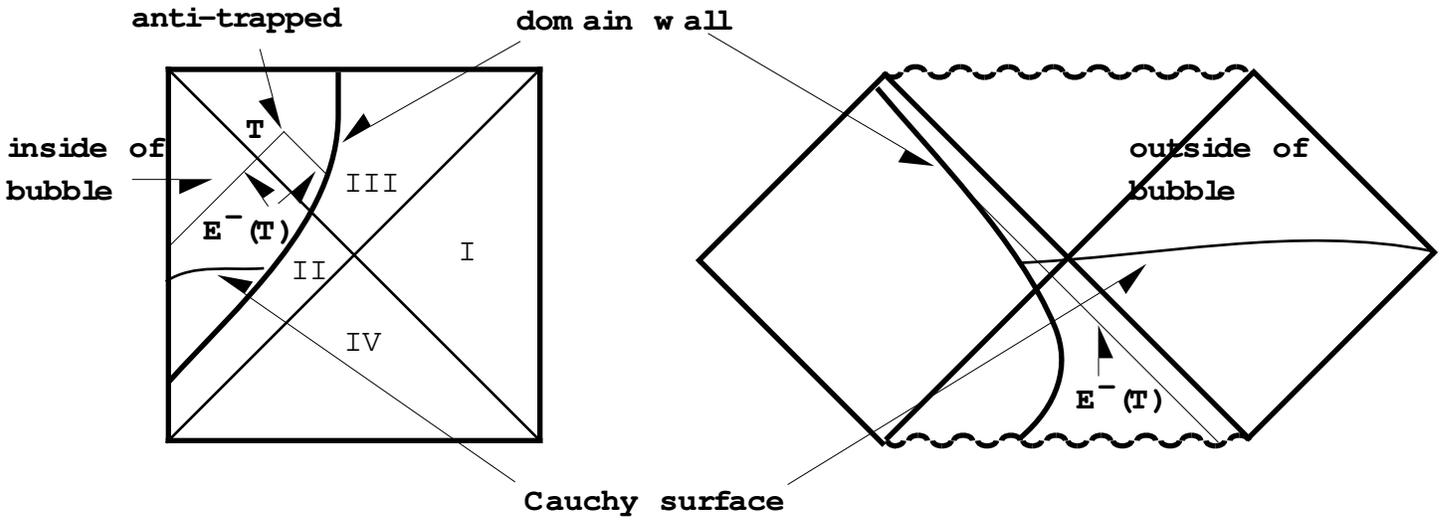
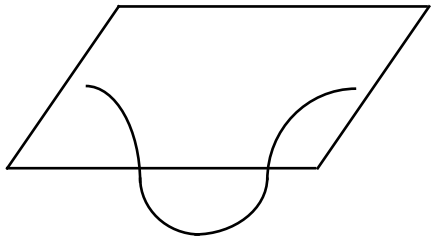
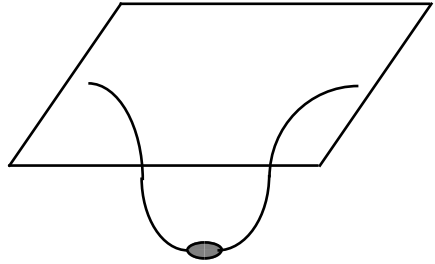
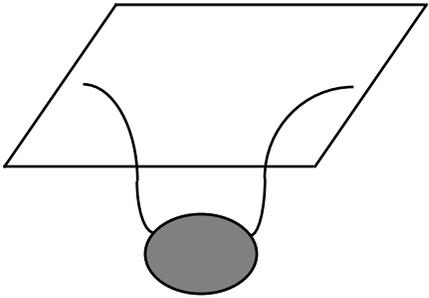
\end{document}